\documentclass[10.5pt,nofootinbib]{revtex4}%
\usepackage{setspace}
\usepackage{doublespace}%
\usepackage{latexsym}
\usepackage{mathptmx}
\usepackage{afterpage}
\usepackage{graphicx}
\usepackage{bm}
\usepackage{amsmath}
\usepackage{amsfonts}
\usepackage{amssymb}%
\begin{document}

\title{First passage time distribution for a random walker on a random forcing energy landscape}
\author{Michael Sheinman$^{1}$, Olivier B\'enichou$^{2}$, Rapha\"el Voituriez$^{2}$ and
Yariv Kafri$^{1}$}
\affiliation{$^{1}$Department of Physics, Technion, Haifa 32000, Israel.}
\affiliation{$^{2}$UMR 7600, Universit\'e Pierre et Marie Curie/CNRS, 4 Place Jussieu,
75255 Paris Cedex 05 France.}
\date{\today}

\begin{abstract}
We present an analytical approximation scheme for the first passage time
distribution on a finite interval of a random walker on a random forcing
energy landscape. The approximation scheme captures the behavior of the
distribution over all timescales in the problem. The results are carefully
checked against numerical simulations.

\end{abstract}
\maketitle

\section{Introduction\label{Introduction}}

The dynamics of a random walker on a one-dimensional random forcing (RF)
energy landscapes has been a subject of much interest over the last three
decades. Part of the interest is due to dynamics in quenched disordered
systems. Examples are the dynamics of a random-field Ising model
\cite{Bru84,Gri84,May84,Nat88} and the motion of dislocations in disordered
crystals \cite{Hlr68}. More recently much interest has been due to many
application in biophysical settings. It seems that in such systems random
forcing energy landscapes are the rule and not the exception. In this context,
the dynamics of random walkers on random forcing energy landscapes have found
applications in the mechanical unzipping of DNA
\cite{LN2000,DCBLNP2003,WLKDNP2005}, translocation of biomolecules through
nanopores \cite{Mel2004} and the dynamics of molecular motors
\cite{Harms1997,Kafri2004,Kafri2005,Hexner2009}. In many of these experiments
the first passage time (FPT) distribution is directly measurable and in some,
such as in the translocation of biomolecules through nanopores, it is the most
direct measurement.

\begin{figure}[ptb]
\begin{center}
\includegraphics[
height=2.5789in,
width=3.4247in
]{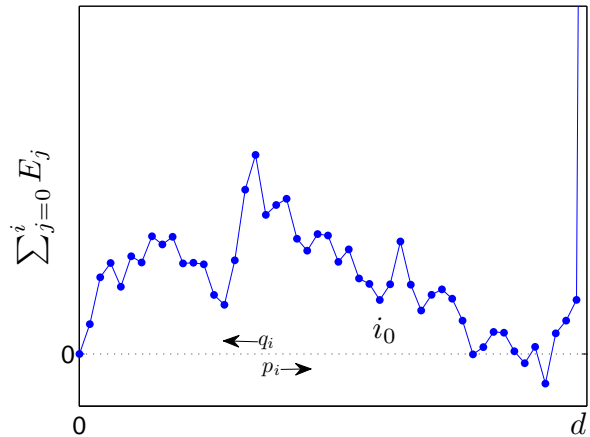}
\end{center}
\caption{An illustration of a RF energy landscape. At site $d+1$ there is a
reflecting boundary. Here $d=50$ and $\mu=0$. The energy difference between
neighboring sites, $E_{i}$, is drawn from the Gaussian distribution.}%
\label{pic2}%
\end{figure}

On a lattice the dynamics of the random walker is defined through the hopping
rates between neighboring sites. We denote the hopping rate from site $i$ to
site $i+1$ by $p_{i}$ and from site $i$ to site $i-1$ by $q_{i}$. To generate
a RF energy landscape the sets of rates $\left\{  p_{i}\right\}  $ and
$\left\{  q_{i}\right\}  $ are drawn randomly from distributions $p\left(
p_{i}\right)  $ and $q\left(  q_{i}\right)  $ respectively (both distributions
are independent of $i$). It is convenient to introduce an energy difference
variables $\left\{  E_{i}\right\}  $ such that $\frac{p_{i-1}}{q_{i}%
}=e^{-E_{i}}$ (measuring energies in units of $k_{B}T$). Thus each realization
of the environment in the RF model can also be drawn as a set of i.i.d. random
variables $\left\{  q_{i}\right\}  $ and $\left\{  E_{i}\right\}  $. Note that
the energy landscape in this model is itself described by a biased random walk
in energy space (see Fig. \ref{pic2} for an illustration).

It is well known \cite{Sin82,Der83a,Kes75,Sol75} that the long-time and
infinite lattice asymptotics of a random walker on such an energy landscape is
rather rich. In particular, the dynamics are controlled by a parameter,
$\mu=\left\vert \widetilde{\mu}\right\vert $, defined through the non-zero
solution of the equation \cite{Der83a,Kes75,Der82b}%
\begin{equation}
\left\langle \left(  \frac{p_{i}}{q_{i+1}}\right)  ^{\widetilde{\mu}%
}\right\rangle =1, \label{23}%
\end{equation}
where the angular brackets represent an average of realizations of disorder
or, equivalently, over the distributions of $\left\{  p_{i}\right\}  $ and
$\left\{  q_{i}\right\}  $. For $\mu>2$ the behavior is similar to that of a
biased random walker on a flat energy landscape. Namely, for large times the
mean position $\left\langle \overline{x}\right\rangle $ and its variance
$\left\langle \overline{\left(  x-\overline{x}\right)  ^{2}}\right\rangle $
both grow linearly in time (here the overline denotes an average over
histories of the system starting from the same initial position). In this
regime both quantities are self averaging. When $0<\mu<2$ the behavior is
anomalous. For $1<\mu<2$ the mean displacement is self averaging and
$\overline{x}\sim\left\langle \overline{x}\right\rangle \sim t$. The
diffusion, however, behaves as $\overline{\left(  x-\overline{x}\right)  ^{2}%
}\sim t^{2/\mu}$ while its average over realizations of disorder behaves as
$\left\langle \overline{\left(  x-\overline{x}\right)  ^{2}}\right\rangle \sim
t^{3-\mu}$. For $0<\mu<1$ both the drift and the diffusion are anomalous.
Asymptotically the velocity, defined through $\underset{t\rightarrow\infty
}{\lim}\left\langle \overline{x}\right\rangle /t$, vanishes. Specifically, the
drift behaves as $\overline{x}\sim\left\langle \overline{x}\right\rangle \sim
t^{\mu}$ (note that although $\overline{x}\ $and $\left\langle \overline
{x}\right\rangle $ have the same scaling $\overline{x}$ is not a
self-averaging quantity \cite{BG90}). In addition $\overline{\left(
x-\overline{x}\right)  ^{2}}\sim t^{2/\mu}$ and its average over the
realizations of disorder behaves as $\left\langle \overline{\left(
x-\overline{x}\right)  ^{2}}\right\rangle \sim t^{3-\mu}$. Finally, when
$\mu=0$ (commonly referred to as Sinai diffusion) $\overline{x}\sim\ln^{2}t$,
$\left\langle \overline{x}\right\rangle =0$, $\overline{x^{2}}\sim\left\langle
\overline{x^{2}}\right\rangle \sim\ln^{4}t$ \cite{Sin82} and $\overline
{\left(  x-\overline{x}\right)  ^{2}}\sim\left\langle \overline{\left(
x-\overline{x}\right)  ^{2}}\right\rangle \sim t^{0}$ \cite{Golosov84}.

In this paper we provide an approximation for the FPT distribution of a random
walker on a RF\ energy landscapes in a finite interval. Namely, we are
interested in the disorder averaged FPT probability density from site
$i_{0}\geq0$ to the origin (site $0$), $F\left(  t\right)  =\left\langle
F_{i_{0}\rightarrow0}\left(  t|\left\{  p_{i},q_{i}\right\}  \right)
\right\rangle $ with reflecting boundary conditions at lattice site $d+1$ with
$d\geq i_{0}$, so that $p_{d}=0$ or $E_{d+1}=\infty$ (see Fig. \ref{pic2}%
)\footnote{To obtain $i_{0}<0$ with boundary condition $q_{d<i_{0}}=0$ results
one simply takes
\begin{align*}
i_{0}  &  \rightarrow-i_{0}\\
q_{i}  &  \rightarrow p_{i}\\
E_{i}  &  \rightarrow-E_{i}\\
d  &  \rightarrow-d.
\end{align*}
}. The results are summarized in Sec. \ref{Summary}. Note that this rather
rich behavior makes it impossible to write the averaged propagator of the
process as a scale invariant function, except in the Sinai's case $\mu=0$.
Therefore, the techniques developed in \cite{Vut1,Vut2} to calculate the
distribution of FPTs for scale invariant processes are not directly
applicable. Our results are compared with numerical calculations and shown to
agree very well. To obtain the approximation we present a "random tilt" (RT)
model, similar in spirit, to that used in \cite{Osh09}. The parameters which
define the RT model are given in terms of the parameters of a random walker on
a RF energy landscape (for simplicity we assume a walker on a lattice)
\textit{with no fitting parameters}. This model naturally exhibits all of the
regimes exhibited by other RF models after a proper disorder average.
Specifically, within the model we obtain an exact result for the Laplace
transform
\begin{equation}
\widetilde{F}\left(  s\right)  =\int_{0}^{\infty}F\left(  t\right)  e^{-st}dt
\end{equation}
of the FPT between two given points with specified boundary condition. The
Laplace transform can be easily inverted numerically.

The paper is organized as follows: in Sec.
\ref{Review of some known results for a random walker on a RF energy landscape}
we present a brief review of relevant known results for a random walker on a
RF energy landscape. In Sec. \ref{Random tilt model} we present and solve the
RT model. In Secs. \ref{Determination of Pr(v)} and
\ref{Determination of x and s} we determine the parameters of the RT model
which are used to approximate the FPT probability density of the RF model. In
Sec. \ref{Comparison} we compare the approximation to numerical results.

\section{Review of some known results for a random walker on a RF energy
landscape\label{Review of some known results for a random walker on a RF energy landscape}%
}

While the exact full FPT distribution, $F\left(  t\right)  $, for which we
provide an approximation, is not known, several related results are known.
Specifically, the mean FPT, defined through $\left\langle \overline
{t}\right\rangle =\int_{0}^{\infty}tF\left(  t\right)  dt$, can be obtained
using the expression for a given realization of disorder (namely, a given set
of $\left\{  E_{i}\right\}  $ and $\left\{  q_{i}\right\}  $) \cite{Mur89}%
\begin{equation}
\overline{t}=\underset{i=1}{\overset{i_{0}}{%
%TCIMACRO{\dsum }%
%BeginExpansion
{\displaystyle\sum}
%EndExpansion
}}\left(  \frac{1}{q_{d}}\underset{j=i_{0}}{\overset{d-1}{%
%TCIMACRO{\dprod }%
%BeginExpansion
{\displaystyle\prod}
%EndExpansion
}}\frac{p_{j}}{q_{j}}+\underset{k=1}{\overset{d-1}{%
%TCIMACRO{\dsum }%
%BeginExpansion
{\displaystyle\sum}
%EndExpansion
}}\frac{1}{q_{k}}\underset{j=i}{\overset{k-1}{%
%TCIMACRO{\dprod }%
%BeginExpansion
{\displaystyle\prod}
%EndExpansion
}}\frac{p_{j}}{q_{j}}\right)  , \label{3}%
\end{equation}
where, as stated above, $\frac{p_{i-1}}{q_{i}}=e^{-E_{i}}$. Therefore, the
disorder average of $\overline{t}$ is
\begin{equation}
\left\langle \overline{t}\right\rangle =\left\langle \frac{1}{q_{i}%
}\right\rangle \frac{\left\langle e^{-E_{i}}\right\rangle ^{d+1}-\left\langle
e^{-E_{i}}\right\rangle ^{d-i_{0}+1}}{\left(  \left\langle e^{-E_{i}%
}\right\rangle -1\right)  ^{2}}-\left\langle \frac{1}{q_{i}}\right\rangle
\frac{i_{0}}{\left\langle e^{-E_{i}}\right\rangle -1}. \label{19}%
\end{equation}
Note that this quantity may be very different from the typical FPT, defined
through $\exp\left\langle \ln\overline{t}\right\rangle $ (see for example
\cite{Dou89b} for a discussion of the Sinai's case $\mu=0$). From Eq.
(\ref{3}) one may see that in the large $d$ limit the leading order of
$\left\langle \ln\overline{t}\right\rangle $ is $\underset{j=1}{\overset
{d}{\sum}}\ln\frac{p_{j}}{q_{j}}=-Ed$, where $E=\left\langle E_{i}%
\right\rangle $ is the average tilt of the RF energy landscape. Therefore, the
typical FPT scales as%
\begin{equation}
\exp\left\langle \ln\overline{t}\right\rangle \sim e^{-Ed}, \label{21}%
\end{equation}
increasing exponentially with $d$ for $E<0$ and remaining constant for $E>0$.

In the limit $d\rightarrow\infty$\ the FPT probability density for $E<0$ is
not normalizable since the probability to never pass $i=0$ is positive. For
the $E>0$ (and $d\rightarrow\infty$) case it is known \cite{BG90,Dou99} that
the mean FPT distribution density scales as $\overline{t}^{-(1+\mu)}$ for
large $\overline{t}$. This implies that the value of $\mu$ may be evaluated by
finding the largest converging moment of the mean FPT for $d\rightarrow\infty
$. Equivalently, $\mu$ is given by the smallest moment that diverges%
\begin{equation}
\mu=\inf\left\{  m:\left\langle \overline{t}^{m}\right\rangle =\infty\right\}
.\label{20}%
\end{equation}
To obtain an approximation for the full FPT distribution, in the next section
we introduce and solve a RT model. The parameters of the RT model are set by
the known properties of the RF model.

\section{The random tilt model\label{Random tilt model}}

\begin{figure}[ptb]
\begin{center}
\includegraphics[width=10cm ]{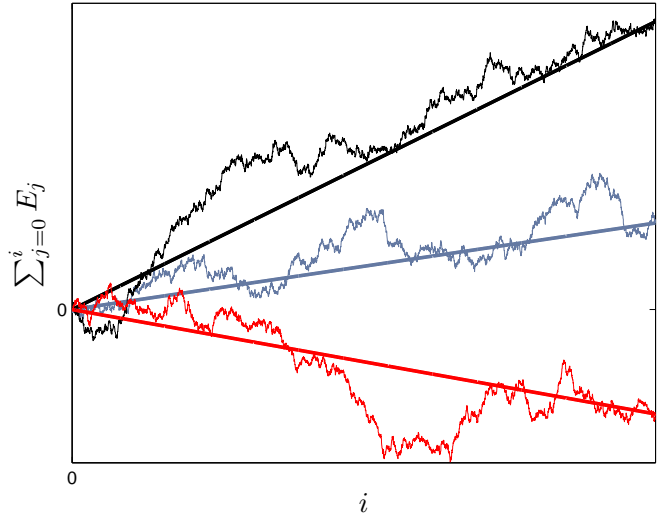}
\end{center}
\caption{On this schematic plot we demonstrate the approach described in
\cite{Osh09} where each specific realization of the disordered is replaced by
constant energy slope.}%
\label{pic1}%
\end{figure}

Recently Oshanin and Redner \cite{Osh09} used an optimal fluctuation method
\cite{Lif88} to study the splitting probability, $P$, of a random walker on a
random forcing energy landscape. The splitting probability is defined as the
probability to reach site $i=d$ before hitting site $i=0$, starting from site
$i=i_{0}$. Within this approach one replaces each specific realization of the
random forcing energy landscape by its average slope, $U$, which is specified
by the energy difference between the origin and the last, $i=d$, site divided
by the total number of the sites (see Fig. \ref{pic1}). The central limit
theorem implies that the probability density of the average energy tilt is
Gaussian:%
\begin{equation}
\Pr\left(  U\right)  =\frac{e^{-d\frac{\left(  U-E\right)  ^{2}}{2\sigma^{2}}%
}}{\sqrt{2\pi\frac{\sigma^{2}}{d}}},
\end{equation}
where $E$, as stated before, is the average energy difference between two
subsequent sites
\begin{equation}
E=\left\langle E_{i}\right\rangle =\left\langle \ln\frac{q_{i}}{p_{i-1}%
}\right\rangle
\end{equation}
and $\sigma^{2}$ is the variance of the energy difference between two
subsequent sites%
\begin{equation}
\sigma^{2}=\left\langle E_{i}^{2}\right\rangle -\left\langle E_{i}%
\right\rangle ^{2}=\left\langle \ln^{2}\frac{q_{i}}{p_{i-1}}\right\rangle
-\left\langle \ln\frac{q_{i}}{p_{i-1}}\right\rangle ^{2}.
\end{equation}
Using the fact that on a constant energy tilt, $U$, the splitting probability
is given by $\frac{1-e^{Ui_{0}}}{1-e^{Ud}}$ \cite{Fel66} the average splitting
probability can be well approximated by \cite{Osh09}%
\begin{equation}
P\simeq\int_{-\infty}^{\infty}\frac{e^{-d\frac{\left(  U-E\right)  ^{2}%
}{2\sigma^{2}}}}{\sqrt{2\pi\frac{\sigma^{2}}{d}}}\frac{1-e^{Ui_{0}}}{1-e^{Ud}%
}dU. \label{5}%
\end{equation}

Inspired by this approach we introduce a random tilt (RT) model. Within the
model the energy landscape of each realization is flat while its tilt is a
random variable. Namely, defining hopping rates to the right and left by $u$
and $v$ respectively, with $\frac{u}{v}=e^{-\varepsilon}$, we take
$\varepsilon$ to be a random variable with a Gaussian probability density%

\begin{equation}
\Pr\left(  \varepsilon\right)  =\frac{1}{\sqrt{2\pi\sigma_{\varepsilon}^{2}%
/d}}e^{-d\frac{\left(  \varepsilon-\left\langle \varepsilon\right\rangle
\right)  ^{2}}{2\sigma_{\varepsilon}^{2}}}%
\end{equation}
with a mean $\left\langle \varepsilon\right\rangle $ and a variance
$\sigma_{\varepsilon}^{2}$. Here $d$, as before, is the size of the system. As
we show below by fixing $\left\langle \varepsilon\right\rangle $,
$\sigma_{\varepsilon}$ and an overall time we can approximate very well the
first passage behavior of a random walker on a RF energy landscape in the
limit $d\rightarrow\infty$ and large $t$. We refer to this as the universal
regime. In this paper, however, we are interested in an approximation over any
time scale when $d$ is finite. In this case the small $t$ and large $t$
behaviors are not universal. As we show a very good approximation can be
achieved by taking the value of $v$ as a random variable with its own
distribution $\Pr\left(  v\right)  $. In Secs. \ref{Determination of x and s}
and \ref{Determination of Pr(v)} we show that, in order to approximate the FPT
probability density of the RF model, the simplest choice one can make is%
\begin{equation}
\Pr\left(  v\right)  =\pi_{1}\delta\left(  v-v_{1}\right)  +\pi_{2}%
\delta\left(  v-v_{2}\right)  +\left(  1-\pi_{1}-\pi_{2}\right)  \delta\left(
v-v_{3}\right)  .
\end{equation}
The values of $\left\langle \varepsilon\right\rangle $, $\sigma_{\varepsilon}%
$, $v_{1}$, $v_{2}$, $v_{3}$, $\pi_{1}$ and $\pi_{2}$ are set by known
properties of the RF model (with no fitting parameters). As stated above, the
results are summarized at the end of the paper. Note that to approximate only
part of the FPT distribution a simpler choice of $\Pr\left(  v\right)  $ can
be made. For example, if one is not interested in the short time behavior one
can chose $\Pr\left(  v\right)  =\delta\left(  v-v_{0}\right)  $ where $v_{0}$
is specified in what follows (see Eq. (\ref{12}) below). If one is not
interested in the long time behavior but wants to capture the short time
behavior one can chose $\Pr\left(  v\right)  =\delta\left(  v-\left\langle
q\right\rangle \right)  $ (see Eq. (\ref{1}) below).

\section{The properties of the RT model}

To approximate the FPT distribution on a RF energy landscape we evaluate the
splitting probability of the RT\ model, $P_{RT}$, the analog of the exponent
$\mu$ of the RT model, denoted by $\mu_{RT}$ and the average FPT, denoted by
$\left\langle \overline{t}\right\rangle _{RT}$. In addition we also calculate
the Laplace transform of the FPT probability density of the RT model,
$\widetilde{F}_{RT}\left(  s\right)  $.

The splitting probability of the RT model can be easily obtained, similar to
Eq. (\ref{5}), and is given by
\begin{equation}
P_{RT}=\int_{-\infty}^{\infty}\frac{e^{-d\frac{\left(  \varepsilon
-\left\langle \varepsilon\right\rangle \right)  ^{2}}{2\sigma_{\varepsilon
}^{2}}}}{\sqrt{2\pi\frac{\sigma_{\varepsilon}^{2}}{d}}}\frac{1-e^{\varepsilon
i_{0}}}{1-e^{\varepsilon d}}d\varepsilon. \label{15}%
\end{equation}
Note that Eq. (\ref{15}) is exact for the RT model.

First we demonstrate that an analog of $\mu$, denoted by $\mu_{RT}$, exists
for the RT\ model when $d\rightarrow\infty$. This shows that indeed the model
can reproduce a qualitative first passage behavior similar to a random walker
on a RF energy landscape. To do this we calculate the scaling behavior of the
moments of the average FPT, $\overline{t}_{\varepsilon}$. For a given
realization of $\varepsilon$ (the equivalent of a random realization of the RF
energy landscape) this scales as%
\begin{equation}
\overline{t}_{\varepsilon}=\frac{1}{v}\frac{\left(  \frac{u}{v}\right)
^{d}-1}{\frac{u}{v}-1}\sim\left\{
\begin{array}
[c]{cc}%
e^{-\varepsilon d} & \varepsilon<0\\
\frac{1}{v}\frac{1}{1-e^{-\varepsilon}} & \varepsilon>0
\end{array}
\right.  . \label{24}%
\end{equation}
Recalling that the probability density of $\varepsilon$ is Gaussian%
\begin{equation}
\Pr\left(  \varepsilon\right)  \sim e^{-d\frac{\left(  \varepsilon
-\left\langle \varepsilon\right\rangle \right)  ^{2}}{2\sigma_{\varepsilon
}^{2}}}\;,
\end{equation}
and in analogy with the RF model (see Eq. (\ref{20})), $\mu_{RT}$ is given by
\begin{equation}
\mu_{RT}=\inf\left\{  m:\left\langle \overline{t_{\varepsilon}}^{m}%
\right\rangle =\infty\right\}  , \label{newmu}%
\end{equation}
where now the angular brackets denote an average over $\varepsilon$ values. In
the $d\rightarrow\infty$ limit, when $\left\langle \overline{t_{\varepsilon}%
}^{m}\right\rangle =\infty$, the average of the $m$'th moment is controlled by
the contributions where $\varepsilon<0$ so that%
\begin{equation}
\left\langle \overline{t}_{\varepsilon}^{m}\right\rangle \sim e^{\left(
-m\left\langle \varepsilon\right\rangle +m^{2}\frac{\sigma_{\varepsilon}^{2}%
}{2}\right)  d}. \label{16}%
\end{equation}
We, therefore, obtains
\begin{equation}
\mu_{RT}=\left\vert \frac{2\left\langle \varepsilon\right\rangle }%
{\sigma_{\varepsilon}^{2}}\right\vert . \label{18}%
\end{equation}
Thus, as stated above, an analog of $\mu$ exists within the RT model after
averaging over realizations of disorder. It, of course, does not exist for
each realization of the model. Note that this expression is identical to that
obtained for a full random forcing model where the random force is drawn from
a Gaussian distribution with a mean $\langle\varepsilon\rangle$ and variance
$\sigma_{\varepsilon}^{2}$ \cite{BG90}. Furthermore, Eq. (\ref{24}) implies
that the scaling behavior of the typical FPT is given by%
\begin{equation}
\exp\left\langle \ln\overline{t}\right\rangle _{RT}\sim e^{-\left\langle
\varepsilon\right\rangle d}, \label{17}%
\end{equation}
increasing exponentially with $d$ for $\left\langle \varepsilon\right\rangle
<0$ and remaining constant when $\left\langle \varepsilon\right\rangle >0$.

Next we consider the mean FPT of the RT model (averaged over realizations of
the tilt of the landscape) to the origin from site $i_{0}$. For a given energy
tilt, $\varepsilon$, and a given left hopping rate, $v$, the thermal averaged
FPT may be obtained using Eq. (\ref{3}) with the substitutions $\left\langle
\frac{1}{q_{i}}\right\rangle =\frac{1}{v}$ and $\left\langle e^{-E_{i}%
}\right\rangle =e^{-\varepsilon}$. Averaging over $\varepsilon$ and $v$ one
obtains
\begin{equation}
\left\langle \overline{t}\right\rangle _{RT}=\left\langle \frac{1}%
{v}\right\rangle I \label{14}%
\end{equation}
where%
\begin{equation}
I=\int_{-\infty}^{\infty}\left[  \frac{e^{-\varepsilon\left(  d+1\right)
}-e^{-\varepsilon\left(  d-i_{0}+1\right)  }}{\left(  e^{-\varepsilon
}-1\right)  ^{2}}-\frac{i_{0}}{e^{-\varepsilon}-1}\right]  \frac
{e^{-d\frac{\left(  \varepsilon-\left\langle \varepsilon\right\rangle \right)
^{2}}{2\sigma_{\varepsilon}^{2}}}}{\sqrt{2\pi\frac{\sigma_{\varepsilon}^{2}%
}{d}}}d\varepsilon\label{13}%
\end{equation}
and%
\begin{equation}
\left\langle \frac{1}{v}\right\rangle =\int_{0}^{\infty}\frac{1}{v}\Pr\left(
v\right)  dv.
\end{equation}

The Laplace transform of the FPT probability density can also be obtained
exactly and is given by%
\begin{equation}
\widetilde{F}_{RT}\left(  s\right)  =\int\int\Pr\left(  \varepsilon\right)
\Pr\left(  v\right)  \Phi\left(  \frac{s}{v},\varepsilon\right)  d\varepsilon
dv, \label{8}%
\end{equation}
where $\Phi\left(  \frac{s}{v},\varepsilon\right)  $ is the FPT probability
density on a flat energy landscape with a tilt $\varepsilon$ and a left
hopping rate $v$. Using standard FPT results \cite{Fel66} one has
\begin{equation}
\Phi\left(  \frac{s}{v},\varepsilon\right)  =\frac{\lambda_{2}^{i_{0}}\left(
\frac{s}{v},\varepsilon\right)  }{2^{i_{0}}}\frac{1-\left(  \frac{\lambda
_{2}\left(  \frac{s}{v},\varepsilon\right)  }{\lambda_{1}\left(  \frac{s}%
{v},\varepsilon\right)  }\right)  ^{d-i_{0}-1}\frac{\left(  \frac{s}%
{v}+1\right)  \lambda_{2}\left(  \frac{s}{v},\varepsilon\right)  -2}{\left(
\frac{s}{v}+1\right)  \lambda_{1}\left(  \frac{s}{v},\varepsilon\right)  -2}%
}{1-\left(  \frac{\lambda_{2}\left(  \frac{s}{v},\varepsilon\right)  }%
{\lambda_{1}\left(  \frac{s}{v},\varepsilon\right)  }\right)  ^{d-1}%
\frac{\left(  \frac{s}{v}+1\right)  \lambda_{2}\left(  \frac{s}{v}%
,\varepsilon\right)  -2}{\left(  \frac{s}{v}+1\right)  \lambda_{1}\left(
\frac{s}{v},\varepsilon\right)  -2}} \label{31}%
\end{equation}
with%
\begin{equation}
\lambda_{1,2}\left(  \frac{s}{v},\varepsilon\right)  =1+\left(  1+\frac{s}%
{v}\right)  e^{\varepsilon}\pm\sqrt{1+2e^{\varepsilon}\left(  \frac{s}%
{v}-1\right)  +e^{2\varepsilon}\left(  \frac{s}{v}+1\right)  ^{2}}. \label{25}%
\end{equation}
Next we use these results to approximate the FPT distribution in the RF\ model.

\section{Approximating the FPT distribution of the RF\ model}

To use the results to approximate the FPT probability density of the RF\ model
here we set the parameters $\left\langle \varepsilon\right\rangle $ and
$\sigma_{\varepsilon}^{2}$ and the probability distribution $\Pr\left(
v\right)  $, such that $\widetilde{F}_{RT}\left(  s\right)  $ yields a good
approximation to $\widetilde{F}\left(  s\right)  $. Namely, we work with the
Laplace transform. The matching is done so the distributions agree in
different ranges of $s$ and can be done by matching between known quantities
of both the RT and the RF models. We first determine the parameters
$\left\langle \varepsilon\right\rangle $ and $\sigma_{\varepsilon}$.

\subsection{Determining $\left\langle \varepsilon\right\rangle $ and
$\sigma_{\varepsilon}$\label{Determination of x and s}}

Here we find expressions for $\left\langle \varepsilon\right\rangle $ and
$\sigma_{\varepsilon}$ by matching the \textit{scaling} properties of the FPT
probability densities. The value of $\left\langle \varepsilon\right\rangle $
may be found by matching the scaling behavior of the typical FPT of the two
models in the large $d$ limit. We showed above (see Eq. (\ref{21})) that in
the RF model the typical FPT grows as $e^{-Ed}$ for $E<0$ and remains constant
while $E>0$ (see Eq. (\ref{24})). The same is true (see Eq. (\ref{17})) for
the RT model when $E$ is replaced by $\left\langle \varepsilon\right\rangle $.
Thus to match the scaling behavior of the typical FPT in both models we set%
\begin{equation}
\left\langle \varepsilon\right\rangle =E. \label{2}%
\end{equation}

The scaling properties in the intermediate $s$ regime for a finite value of
$d$ can be expected to behave identically to the \textit{small} $s$ behavior
of the $d\rightarrow\infty$ limit. Since the scaling behavior is different for
negative and positive average tilts we separate our discussion to two cases.

\subsubsection{The case $E<0$}

In this case the splitting probability, $P$, remains finite in the
$d\rightarrow\infty$ limit. Thus, the small $s$ behavior of the FPT
probability density in the $d\rightarrow\infty$ limit is $1-P+O\left(
s\right)  $. To match the intermediate $s$ behavior of FPT probability density
of RF and RT models we match the splitting probabilities (i.e. the leading,
$O\left(  s^{0}\right)  $, order of the FPT probability densities) of these
two models.

Comparing (\ref{5}) with (\ref{15}) and using Eq. (\ref{2}) yields
\begin{equation}
\sigma_{\varepsilon}=\sigma. \label{22}%
\end{equation}
Note that both the average tilt and the variance in the RF and RT models are
equal in this regime.

\subsubsection{The case $E>0$}

As studied in Sec. \ref{Introduction}, in this case the mean FPT distribution
for $d\rightarrow\infty$ in the small $s$ region scales as $1-O\left(  s^{\mu
}\right)  $ for $\mu<1$ (or, equivalently in time as $\overline{t}^{-\left(
1+\mu\right)  }$). Therefore, we first match $\mu_{RT}$ of the RT model with
$\mu$ of the RF model. Namely, we set $\mu_{RT}=\mu$. Using Eq. (\ref{18})
this yields
\begin{equation}
\frac{2\left\langle \varepsilon\right\rangle }{\sigma_{\varepsilon}^{2}}=\mu.
\label{4}%
\end{equation}
Solving Eqs. (\ref{2}) and (\ref{4}) we get%
\begin{equation}
\sigma_{\varepsilon}=\sqrt{\frac{2E}{\mu}}. \label{7}%
\end{equation}
Note when the distribution of $E_{i}$ is Gaussian equations (\ref{22}) and
(\ref{7}) become identical. Namely, both the average tile and the variance of
the RF and RT models are equivalent. When the distribution is not Gaussian the
average tilt is the same in both models but the variances can be different. In
this regime it is important to match the precise value of $\mu$ rather than
the splitting probability as in the $E<0$ case.

\subsection{Determination of $\Pr\left(  v\right)  $%
\label{Determination of Pr(v)}}

In this section we choose the distribution $\Pr\left(  v\right)  $ by matching
the FPT probability density of the RT model to that of the original RF system
in the small and large $s$ limits. We show that for each limit one may choose
$v$ to be a constant. However, this constant depends on the limit. Thus, to
match the full $s$ behavior $v$ cannot be a constant and, as we show below,
the simplest choice for $\Pr\left(  v\right)  $ is
\begin{equation}
\Pr\left(  v\right)  =\pi_{1}\delta\left(  v-v_{1}\right)  +\pi_{2}%
\delta\left(  v-v_{2}\right)  +\left(  1-\pi_{1}-\pi_{2}\right)  \delta\left(
v-v_{3}\right)
\end{equation}
with $v_{1,2,3}$ are set according to relevant time scales in the problem and
$\pi_{1,2}$ are determined by the matching procedure, described below.

\subsubsection{The small $s$ approximation}

The $s\rightarrow0$ behavior of the FPT probability densities of the RF and
the RT models is%
\begin{equation}
\widetilde{F}\left(  s\rightarrow0\right)  =1-\left\langle \overline
{t}\right\rangle s
\end{equation}
and%
\begin{equation}
\widetilde{F}_{RT}\left(  s\rightarrow0\right)  =1-\left\langle \overline
{t}\right\rangle _{RT}s,
\end{equation}
respectively. Therefore, to match $\widetilde{F}\left(  s\rightarrow0\right)
$ and $\widetilde{F}_{RT}\left(  s\rightarrow0\right)  $ one has to match the
mean FPTs of both models. Demanding $\left\langle \overline{t}\right\rangle
_{RT}=\left\langle \overline{t}\right\rangle $ and using Eq. (\ref{14}) one
has%
\begin{equation}
\left\langle \frac{1}{v}\right\rangle =\frac{\left\langle \overline
{t}\right\rangle }{I} \label{12}%
\end{equation}
where $I$ is defined in Eq. (\ref{13}) and $\left\langle \overline
{t}\right\rangle $ is given by Eq. (\ref{19}).

Below we use this fact to match the behavior of the FPT distribution of the RF
model over the whole $s$ range. Note however, that if one is not interested in
the short time behavior the procedure is simplified. One can then choose
$\Pr\left(  v\right)  =\delta\left(  v-v_{1}\right)  $ where $v_{1}=\frac
{I}{\left\langle \overline{t}\right\rangle }$. Then, using Eq. (\ref{8}), the
Laplace transform of the FPT probability density is given by%
\begin{equation}
\widetilde{F}_{RT}\left(  s\right)  =\int_{-\infty}^{\infty}\frac
{e^{-d\frac{\left(  \varepsilon-\left\langle \varepsilon\right\rangle \right)
^{2}}{2\sigma_{\varepsilon}^{2}}}}{\sqrt{2\pi\frac{\sigma_{\varepsilon}^{2}%
}{d}}}\Phi\left(  \frac{s}{v_{1}},\varepsilon\right)  d\varepsilon.
\label{smalls}%
\end{equation}

\subsubsection{The large $s$ approximation}

The large $s$ limit corresponds to the small $t$ limit. This regime of the FPT
distribution is controlled by walks which hop only to the left before reaching
the origin. Therefore, in the $s\rightarrow\infty$ limit the behavior of the
FPT probability densities of the RF and the RT models is%
\begin{equation}
\widetilde{F}\left(  s\rightarrow\infty\right)  =\left\langle q\right\rangle
^{i_{0}}%
\end{equation}
and%
\begin{equation}
\widetilde{F}_{RT}\left(  s\rightarrow\infty\right)  =\left\langle v^{i_{0}%
}\right\rangle ,
\end{equation}
respectively. Thus, to match the large $s$ behavior of $\widetilde{F}%
_{RT}\left(  s\right)  $ and $\widetilde{F}\left(  s\right)  $ we demand%
\begin{equation}
\left\langle v^{i_{0}}\right\rangle =\left\langle q\right\rangle ^{i_{0}}.
\label{1}%
\end{equation}

Next we match the behaviors in the RT and RF models over the whole range of
$s$. However, we comment that if one is not interested in the approximate FPT
distribution in the long time limit one can choose $\Pr\left(  v\right)
=\delta\left(  v-v_{2}\right)  $ where $v_{2}=\left\langle q\right\rangle $.
Then, using Eq. (\ref{8}), the Laplace transform of the FPT probability
density is given by%
\begin{equation}
\widetilde{F}_{RT}\left(  s\right)  =\int_{-\infty}^{\infty}\frac
{e^{-d\frac{\left(  \varepsilon-\left\langle \varepsilon\right\rangle \right)
^{2}}{2\sigma_{\varepsilon}^{2}}}}{\sqrt{2\pi\frac{\sigma_{\varepsilon}^{2}%
}{d}}}\Phi\left(  \frac{s}{v_{2}},\varepsilon\right)  d\varepsilon.
\label{larges}%
\end{equation}

\subsubsection{An approximation for the whole $s$ range}

To match both the small and large $s$ limit one should supply $\Pr\left(
v\right)  $ such that Eqs. (\ref{12}) and (\ref{1}) are satisfied. However,
these equations do not determine $\Pr\left(  v\right)  $ uniquely and there is
a lot of freedom. As stated above, the simplest choice that satisfies these
equations is\footnote{Other, simple choices lead to nonlinear equations. For
example, using $\Pr\left(  v\right)  =\frac{1}{2}\delta\left(  v-v_{1}\right)
+\frac{1}{2}\delta\left(  v-v_{2}\right)  $, Eqs. (\ref{12}) and (\ref{1})
lead to $\frac{1}{v_{1}}+\frac{1}{v_{2}}=2\frac{\left\langle \overline
{t}\right\rangle }{I}$ and $v_{1}^{i_{0}}+v_{2}^{i_{0}}=2\left\langle
q\right\rangle ^{i_{0}}$.}%
\begin{equation}
\Pr\left(  v\right)  =\pi_{1}\delta\left(  v-v_{1}\right)  +\pi_{2}%
\delta\left(  v-v_{2}\right)  +\left(  1-\pi_{1}-\pi_{2}\right)  \delta\left(
v-v_{3}\right)  . \label{9}%
\end{equation}
However, since one has to satisfy only two equations ((\ref{12}) and
(\ref{1})) there is still a lot of freedom with five free parameters.
Therefore, first we set $v_{1,2,3}$ such that they represent relevant time
scales in the problem. Eq. (\ref{12}) suggests%
\begin{equation}
\frac{1}{v_{1}}=\frac{\left\langle \overline{t}\right\rangle }{I}, \label{28}%
\end{equation}
while Eq. (\ref{1}) suggests
\begin{equation}
v_{2}=\left\langle q\right\rangle . \label{29}%
\end{equation}
The third time scale is chosen to represent the average time of a hop to the
left%
\begin{equation}
v_{3}=\left\langle \frac{1}{q}\right\rangle ^{-1}. \label{30}%
\end{equation}
Given these three quantities, $v_{1,2,3}$, Eqs. (\ref{12}) and (\ref{1}) give%
\begin{equation}
\pi_{1}v_{1}^{i_{0}}+\pi_{2}v_{2}^{i_{0}}+\left(  1-\pi_{1}-\pi_{2}\right)
v_{3}^{i_{0}}=\left\langle \,q\right\rangle ^{i_{0}}%
\end{equation}
and%
\begin{equation}
\frac{\pi_{1}}{v_{1}}+\frac{\pi_{2}}{v_{2}}+\frac{1-\pi_{1}-\pi_{2}}{v_{3}%
}=\frac{\left\langle \overline{t}\right\rangle }{I}.
\end{equation}
The solution for $\pi_{1,2}$ is%
\begin{equation}
\pi_{1}=\frac{v_{1}\left\langle \overline{t}\right\rangle }{I}\frac{v_{2}%
v_{3}\left(  v_{2}^{i_{0}}-v_{3}^{i_{0}}\right)  +\frac{I\left\langle
q\right\rangle ^{i_{0}}}{\left\langle \overline{t}\right\rangle }\left(
v_{2}-v_{3}\right)  +\frac{I}{\left\langle \overline{t}\right\rangle }\left(
v_{3}^{i_{0}+1}-v_{2}^{i_{0}+1}\right)  }{v_{2}v_{3}\left(  v_{2}^{i_{0}%
}-v_{3}^{i_{0}}\right)  +v_{1}^{i_{0}+1}\left(  v_{2}-v_{3}\right)
+v_{1}\left(  v_{3}^{i_{0}+1}-v_{2}^{i_{0}+1}\right)  } \label{26}%
\end{equation}
and%
\begin{equation}
\pi_{2}=-\frac{v_{2}\left\langle \overline{t}\right\rangle }{I}\frac
{v_{1}v_{3}\left(  v_{1}^{i_{0}}-v_{3}^{i_{0}}\right)  +\frac{I\left\langle
q\right\rangle ^{i_{0}}}{\left\langle \overline{t}\right\rangle }\left(
v_{1}-v_{3}\right)  +\frac{I}{\left\langle \overline{t}\right\rangle }\left(
v_{3}^{i_{0}+1}-v_{1}^{i_{0}+1}\right)  }{v_{2}v_{3}\left(  v_{2}^{i_{0}%
}-v_{3}^{i_{0}}\right)  +v_{1}^{i_{0}+1}\left(  v_{2}-v_{3}\right)
+v_{1}\left(  v_{3}^{i_{0}+1}-v_{2}^{i_{0}+1}\right)  }. \label{27}%
\end{equation}
Using these quantities, the probability density (\ref{9}) for $v$ and Eq.
(\ref{8}), the Laplace transform of the FPT probability density is given by%
\begin{equation}
\widetilde{F}_{RT}\left(  s\right)  =\int_{-\infty}^{\infty}\frac
{e^{-d\frac{\left(  \varepsilon-\left\langle \varepsilon\right\rangle \right)
^{2}}{2\sigma_{\varepsilon}^{2}}}}{\sqrt{2\pi\frac{\sigma_{\varepsilon}^{2}%
}{d}}}\left[  \pi_{1}\Phi\left(  \frac{s}{v_{1}},\varepsilon\right)  +\pi
_{2}\Phi\left(  \frac{s}{v_{2}},\varepsilon\right)  +\left(  1-\pi_{2}-\pi
_{3}\right)  \Phi\left(  \frac{s}{v_{3}},\varepsilon\right)  \right]
d\varepsilon. \label{10}%
\end{equation}
In the next section we compare the obtained result to the numerical
simulation. This expression with the above chosen values of parameters is the
main result of our paper. Note that while quite a few parameters have to be
set there are no free fitting parameters. As we show the agreement is very good.

\subsection{FPT probability density and comparison to numerical
result\label{Comparison}}

In this section we compare between Eq. (\ref{10}) and numerical simulations.
Namely we compare the Laplace transform of the FPT probability, density,
$\widetilde{F}_{RT}\left(  s\right)  $, and the survival probability%
\begin{equation}
F_{S}\left(  t\right)  =1-\int_{0}^{t}F\left(  t^{\prime}\right)  dt^{\prime}%
\end{equation}
of the RF and RT models. We checked the results for RF models with both
Bernoulli and Gaussian disorders.

\begin{figure}[ptb]
\begin{center}
\includegraphics[
width=12cm]{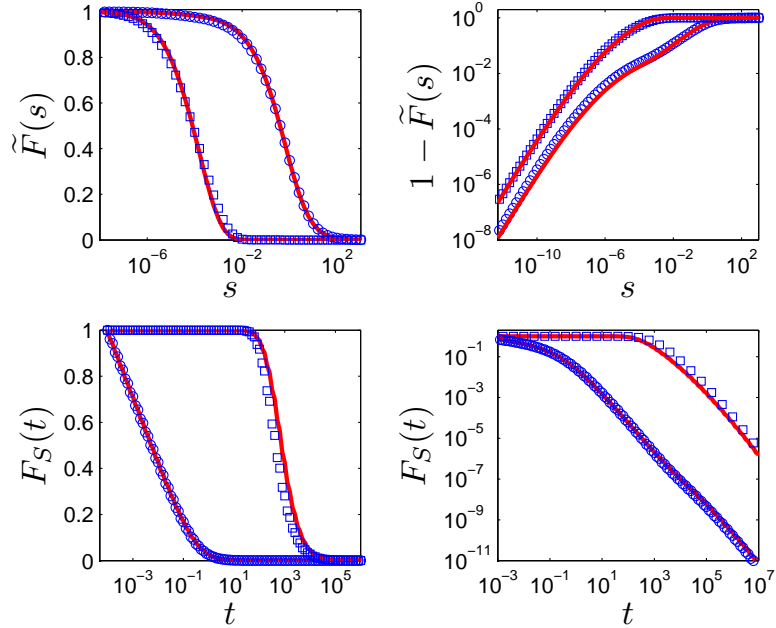}
\end{center}
\caption{In this graph a comparison between the analytic approximation
(\ref{10}) and the numerical results for a Bernoulli disorder is shown in
Laplace (top) and time (bottom) spaces. The parameters for these plots are:
$r=0.5$, $\varepsilon_{1}=-0.4$, $\varepsilon_{2}=0.4$ and $d=100$, such that
$\mu=0$. Circles represent the choice $i_{0}=1$ while squares represent the
choice $i_{0}=50$ cases on both, left and right, figures. The red lines are
analytic approximations based on Eq. (\ref{10}).}%
\label{g1}%
\end{figure}

\begin{figure}[ptb]
\begin{center}
\includegraphics[
width=12cm]{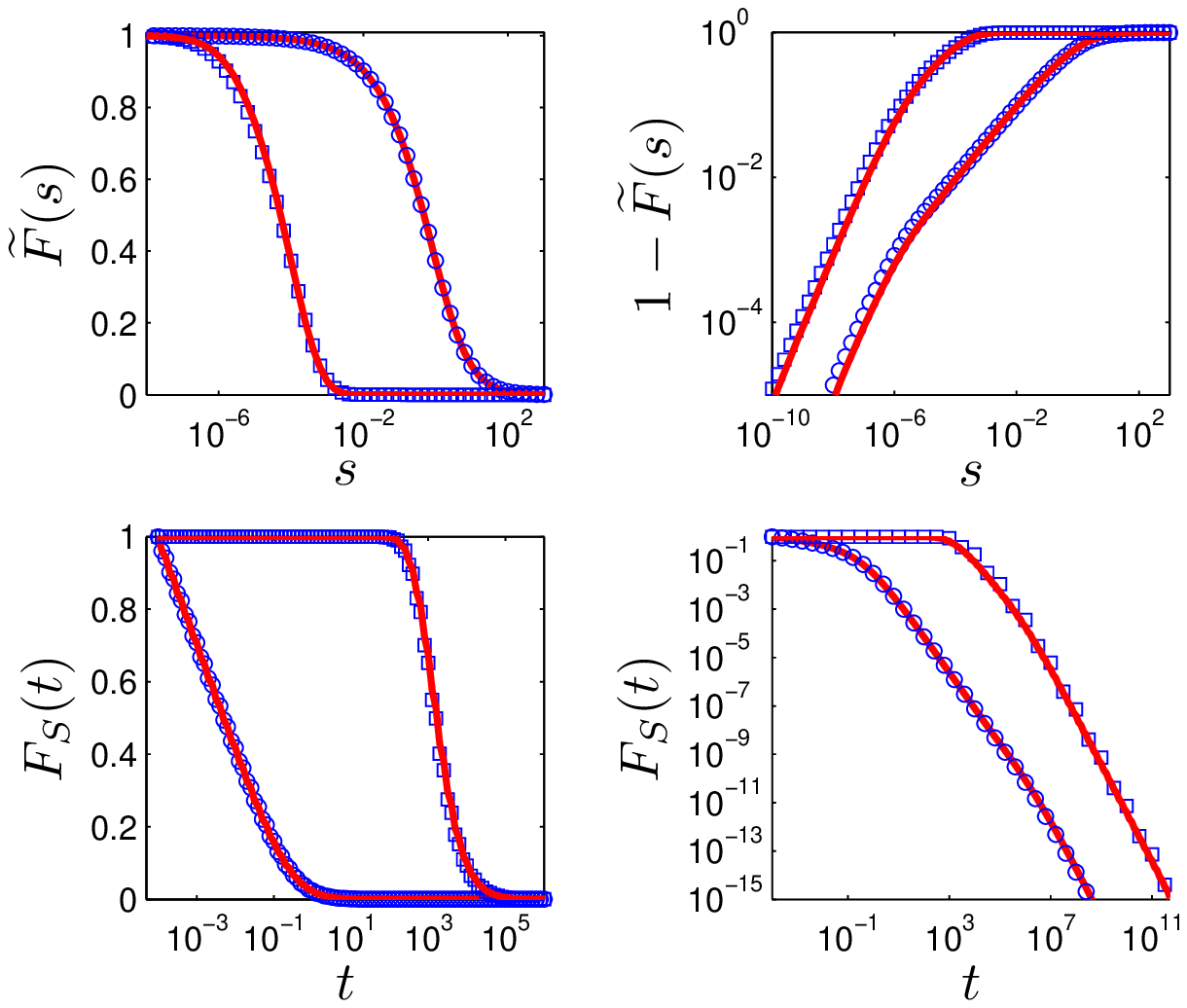}
\end{center}
\caption{In this graph a comparison between the analytic approximation
(\ref{10}) and the numerical results for a Bernoulli disorder is shown in
Laplace (top) and time (bottom) spaces. The parameters for these plots are:
$r=0.49$, $\varepsilon_{1}=-0.1$, $\varepsilon_{2}=0.1$ and $d=550$, such that
$\mu=0.4$. Circles represent the choice $i_{0}=1$ while squares represent the
choice $i_{0}=100$ cases on both, left and right, figures. The red lines are
analytic approximations based on Eq. (\ref{10}).}%
\label{g2}%
\end{figure}

\begin{figure}[ptb]
\begin{center}
\includegraphics[
width=12cm]{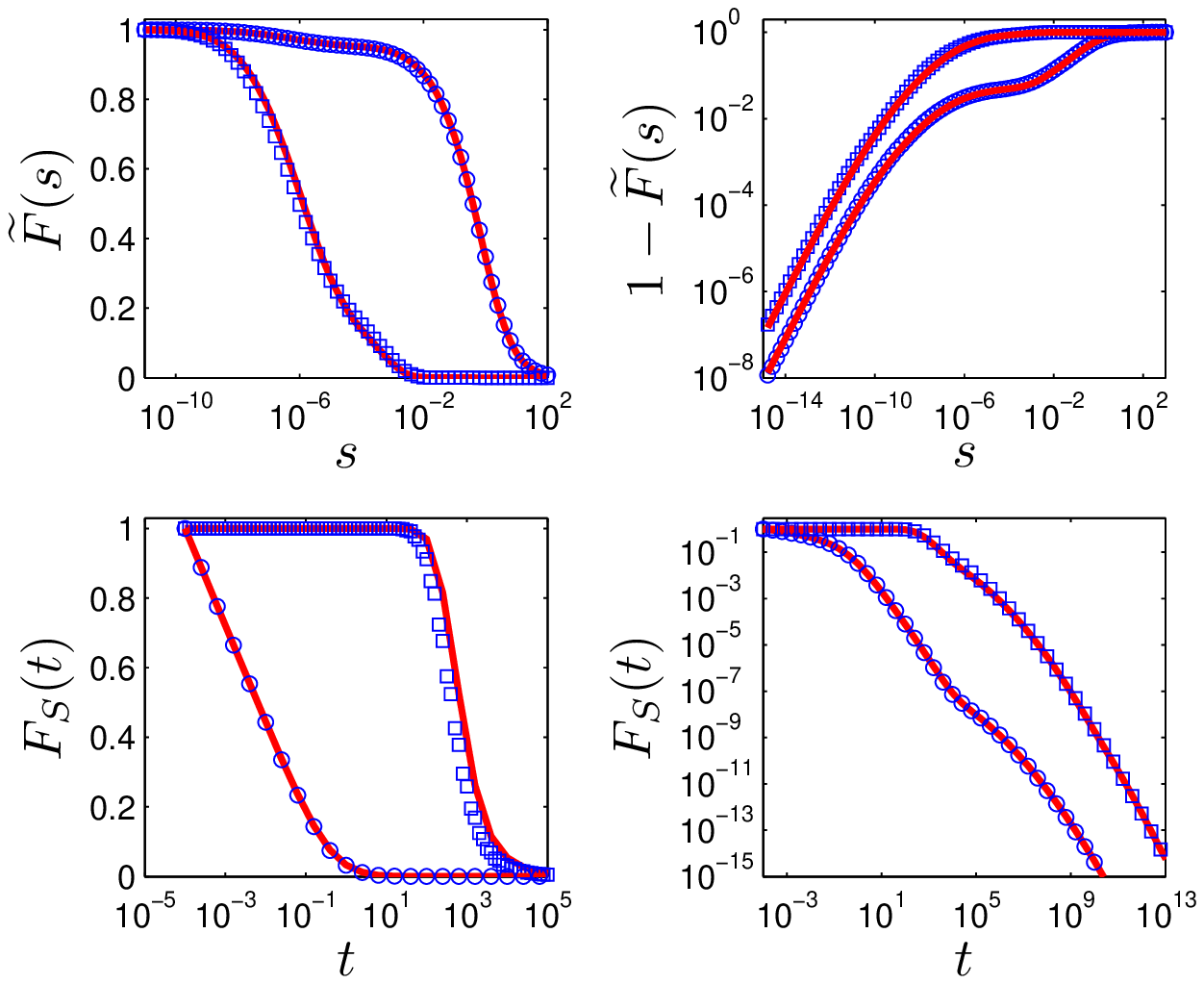}
\end{center}
\caption{In this graph a comparison between the analytic approximation
(\ref{10}) and the numerical results for a Bernoulli disorder is shown in
Laplace (top) and time (bottom) spaces. The parameters for these plots are:
$r=0.58$, $\varepsilon_{1}=-0.3$, $\varepsilon_{2}=0.3$ and $d=150$, such that
$\mu=-1.08$. Circles represent the choice $i_{0}=1$ while squares represent
the choice $i_{0}=50$ cases on both, left and right, figures. The red lines
are analytic approximations based on Eq. (\ref{10}).}%
\label{g3}%
\end{figure}

For the RF system with Bernoulli disorder the energy difference on each site,
$E_{i}$, is drawn from the distribution:
\begin{equation}
\Pr\left(  E_{i}\right)  =\left\{
\begin{array}
[c]{cc}%
r & E_{i}=\varepsilon_{1}\\
1-r & E_{i}=\varepsilon_{2}%
\end{array}
\right.  .
\end{equation}
On Figs. \ref{g1},\ref{g2},\ref{g3} one may see the comparison between Eq.
(\ref{10}) and the numerical calculation in different parameter ranges. As can
be seen the agreement is very satisfying.

\begin{figure}[ptb]
\begin{center}
\includegraphics[
width=12cm]{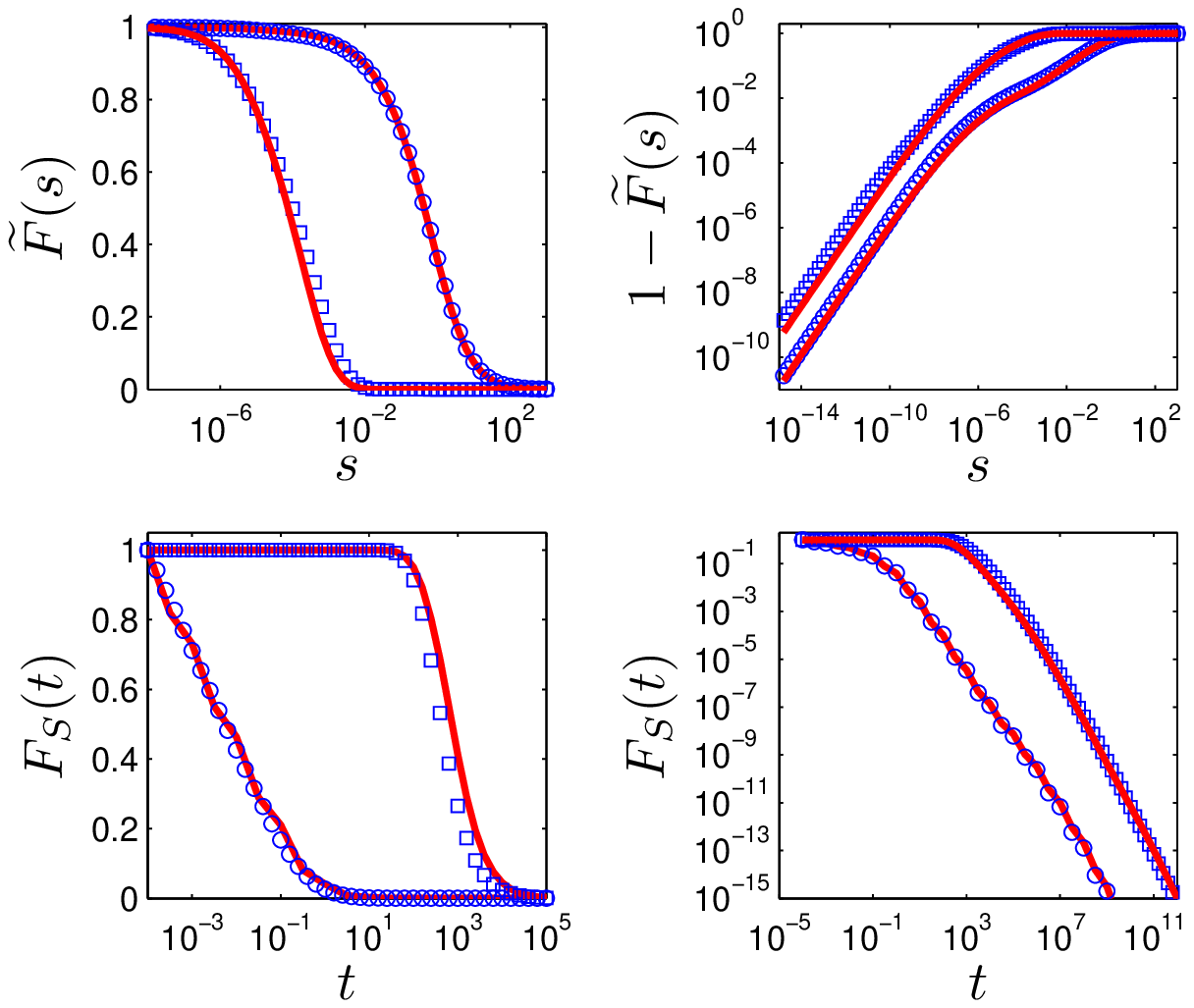}
\end{center}
\caption{In this graph a comparison between the analytic approximation
(\ref{10}) and the numerical results for a Gaussian disorder is shown in
Laplace (top) and time (bottom) spaces. The parameters for these plots are:
$E=0$, $\sigma=0.3$ and $d=150$, such that $\mu=0$. Circles represent the
choice $i_{0}=1$ while squares represent the choice $i_{0}=50$ cases on both,
left and right, figures. The red lines are analytic approximations based on
Eq. (\ref{10}).}%
\label{g4}%
\end{figure}

\begin{figure}[ptb]
\begin{center}
\includegraphics[
width=12cm]{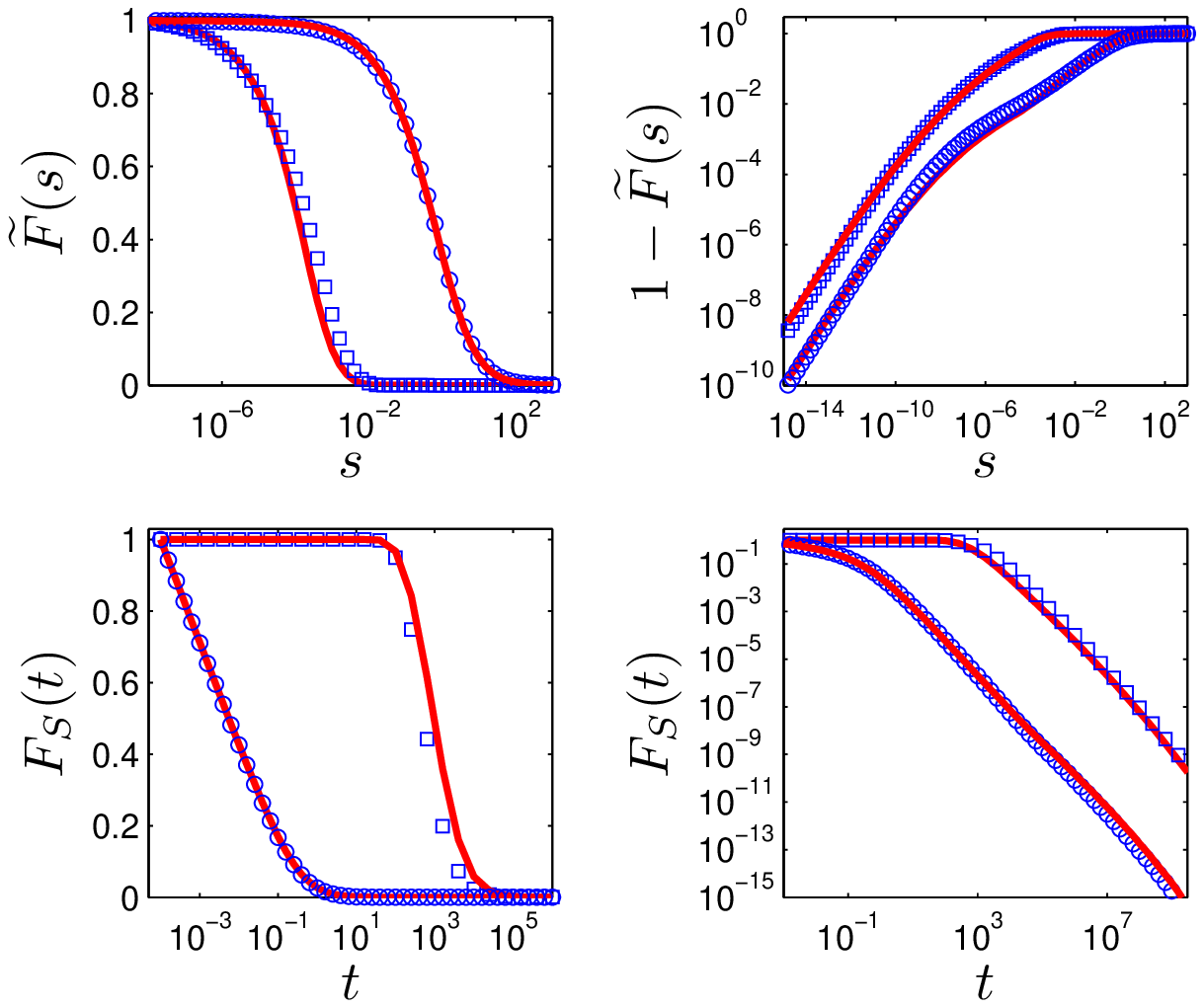}
\end{center}
\caption{In this graph a comparison between the analytic approximation
(\ref{10}) and the numerical results for a Gaussian disorder is shown in
Laplace (top) and time (bottom) spaces. The parameters for these plots are:
$E=0.01$, $\sigma=0.3$ and $d=250$, such that $\mu=0.2222$ Circles represent
the choice $i_{0}=1$ while squares represent the choice $i_{0}=50$ cases on
both, left and right, figures. The red lines are analytic approximations based
on Eq. (\ref{10}).}%
\label{g5}%
\end{figure}

\begin{figure}[ptb]
\begin{center}
\includegraphics[
width=12cm]{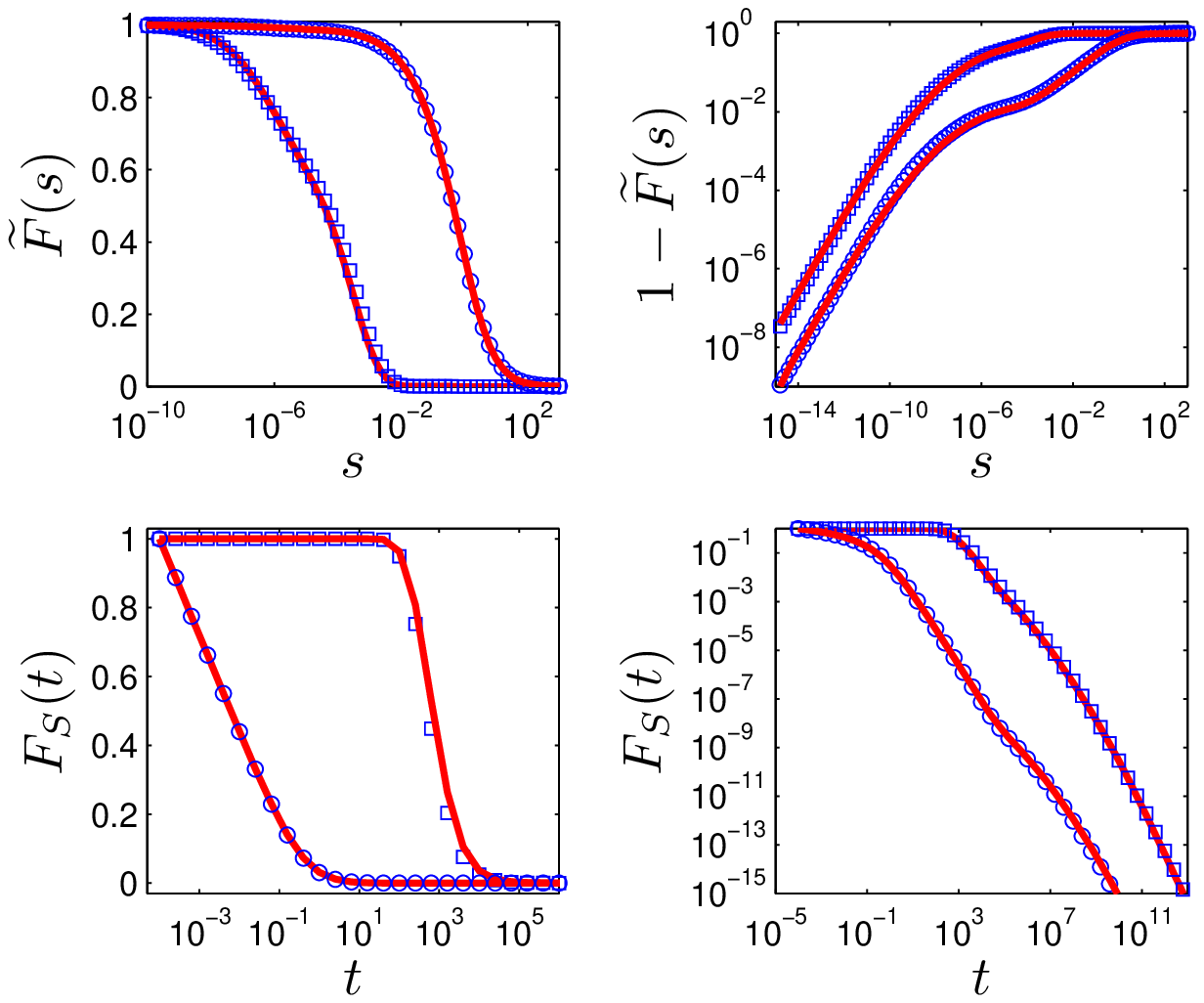}
\end{center}
\caption{In this graph a comparison between the analytic approximation
(\ref{10}) and the numerical results for a Gaussian disorder is shown in
Laplace (top) and time (bottom) spaces. The parameters for these plots are:
$E=-0.01$, $\sigma=0.2$ and $d=350$, such that $\mu=-0.5$ Circles represent
the choice $i_{0}=1$ while squares represent the choice $i_{0}=50$ cases on
both, left and right, figures. The red lines are analytic approximations based
on Eq. (\ref{10}).}%
\label{g6}%
\end{figure}

For a RF model with a Gaussian disorder the energy difference between sites,
$E_{i}$, is drawn from a normal distribution with a mean $E$ and a variance
$\sigma^{2}$:%
\begin{equation}
\Pr\left(  E_{i}\right)  =\frac{e^{-d\frac{\left(  E_{i}-E\right)  ^{2}%
}{2\sigma^{2}}}}{\sqrt{2\pi\frac{\sigma^{2}}{d}}}.
\end{equation}
On Figs. \ref{g4},\ref{g5},\ref{g6} we show a comparison between Eq.
(\ref{10}) and numerical calculation in different parameter ranges. Again, the
results of the approximation are very good.

\section{Summary\label{Summary}}

In this paper we presented a random tilt model and solved it analytically. We
showed that the model can be used to obtain an approximate expression for the
FPT distribution of a random walker on a random forcing energy landscape. To
do this several parameters of the RT model have to be set as a function of the
RF model parameters. As we showed, this can be done with no free fitting
parameters. For convenience we summarize the results below.

The approximation for the random walker's Laplace transformed FPT probability
density from site $i_{0}>0$ to the origin on a random forcing energy landscape
with i.i.d. random variables $\left\{  p_{i}\right\}  $ and $\left\{
q_{i}\right\}  $ is:%

\[
\widetilde{F}\left(  s\right)  =\int_{-\infty}^{\infty}\frac{e^{-d\frac
{\left(  \varepsilon-\left\langle \varepsilon\right\rangle \right)  ^{2}%
}{2\sigma_{\varepsilon}^{2}}}}{\sqrt{2\pi\frac{\sigma_{\varepsilon}^{2}}{d}}%
}\left[  \pi_{1}\Phi\left(  \frac{s}{v_{1}},\varepsilon\right)  +\pi_{2}%
\Phi\left(  \frac{s}{v_{2}},\varepsilon\right)  +\left(  1-\pi_{2}-\pi
_{3}\right)  \Phi\left(  \frac{s}{v_{3}},\varepsilon\right)  \right]
d\varepsilon\text{ (Eq. (\ref{10})),}%
\]
where%

\[
\Phi\left(  \frac{s}{v},\varepsilon\right)  =\frac{\lambda_{2}^{i_{0}}\left(
\frac{s}{v},\varepsilon\right)  }{2^{i_{0}}}\frac{1-\left(  \frac{\lambda
_{2}\left(  \frac{s}{v},\varepsilon\right)  }{\lambda_{1}\left(  \frac{s}%
{v},\varepsilon\right)  }\right)  ^{d-i_{0}-1}\frac{\left(  \frac{s}%
{v}+1\right)  \lambda_{2}\left(  \frac{s}{v},\varepsilon\right)  -2}{\left(
\frac{s}{v}+1\right)  \lambda_{1}\left(  \frac{s}{v},\varepsilon\right)  -2}%
}{1-\left(  \frac{\lambda_{2}\left(  \frac{s}{v},\varepsilon\right)  }%
{\lambda_{1}\left(  \frac{s}{v},\varepsilon\right)  }\right)  ^{d-1}%
\frac{\left(  \frac{s}{v}+1\right)  \lambda_{2}\left(  \frac{s}{v}%
,\varepsilon\right)  -2}{\left(  \frac{s}{v}+1\right)  \lambda_{1}\left(
\frac{s}{v},\varepsilon\right)  -2}}\text{ (Eq. (\ref{31})),}%
\]%
\[
\lambda_{1,2}\left(  \frac{s}{v},\varepsilon\right)  =1+\left(  1+\frac{s}%
{v}\right)  e^{\varepsilon}\pm\sqrt{1+2e^{\varepsilon}\left(  \frac{s}%
{v}-1\right)  +e^{2\varepsilon}\left(  \frac{s}{v}+1\right)  ^{2}}\text{ (Eq.
(\ref{25})),}%
\]%
\[
\left\langle \overline{t}\right\rangle =\left\langle \frac{1}{q_{i}%
}\right\rangle \frac{\left\langle e^{-E_{i}}\right\rangle ^{d+1}-\left\langle
e^{-E_{i}}\right\rangle ^{d-i_{0}+1}}{\left(  \left\langle e^{-E_{i}%
}\right\rangle -1\right)  ^{2}}-\left\langle \frac{1}{q_{i}}\right\rangle
\frac{i_{0}}{\left\langle e^{-E_{i}}\right\rangle -1}\text{(Eq. (\ref{19})),}%
\]%
\[
I=\int_{-\infty}^{\infty}\left[  \frac{e^{-\varepsilon\left(  d+1\right)
}-e^{-\varepsilon\left(  d-i_{0}+1\right)  }}{\left(  e^{-\varepsilon
}-1\right)  ^{2}}-\frac{i_{0}}{e^{-\varepsilon}-1}\right]  \frac
{e^{-d\frac{\left(  \varepsilon-\left\langle \varepsilon\right\rangle \right)
^{2}}{2\sigma_{\varepsilon}^{2}}}}{\sqrt{2\pi\frac{\sigma_{\varepsilon}^{2}%
}{d}}}d\varepsilon\text{ (Eq. (\ref{13})),}%
\]%
\[
v_{1}=\frac{I}{\left\langle \overline{t}\right\rangle }\text{ (Eq.
(\ref{28})),}%
\]%
\[
v_{2}=\left\langle q_{i}\right\rangle \text{ (Eq. (\ref{29}))},
\]%
\[
v_{3}=\left\langle \frac{1}{q_{i}}\right\rangle ^{-1}\text{ (Eq. (\ref{30})),}%
\]%
\[
\pi_{1}=\frac{v_{1}\left\langle \overline{t}\right\rangle }{I}\frac{v_{2}%
v_{3}\left(  v_{2}^{i_{0}}-v_{3}^{i_{0}}\right)  +\frac{I\left\langle
q\right\rangle ^{i_{0}}}{\left\langle \overline{t}\right\rangle }\left(
v_{2}-v_{3}\right)  +\frac{I}{\left\langle \overline{t}\right\rangle }\left(
v_{3}^{i_{0}+1}-v_{2}^{i_{0}+1}\right)  }{v_{2}v_{3}\left(  v_{2}^{i_{0}%
}-v_{3}^{i_{0}}\right)  +v_{1}^{i_{0}+1}\left(  v_{2}-v_{3}\right)
+v_{1}\left(  v_{3}^{i_{0}+1}-v_{2}^{i_{0}+1}\right)  }\text{ (Eq.
(\ref{26})),}%
\]%
\[
\pi_{2}=-\frac{v_{2}\left\langle \overline{t}\right\rangle }{I}\frac
{v_{1}v_{3}\left(  v_{1}^{i_{0}}-v_{3}^{i_{0}}\right)  +\frac{I\left\langle
q\right\rangle ^{i_{0}}}{\left\langle \overline{t}\right\rangle }\left(
v_{1}-v_{3}\right)  +\frac{I}{\left\langle \overline{t}\right\rangle }\left(
v_{3}^{i_{0}+1}-v_{1}^{i_{0}+1}\right)  }{v_{2}v_{3}\left(  v_{2}^{i_{0}%
}-v_{3}^{i_{0}}\right)  +v_{1}^{i_{0}+1}\left(  v_{2}-v_{3}\right)
+v_{1}\left(  v_{3}^{i_{0}+1}-v_{2}^{i_{0}+1}\right)  }\text{ (Eq.
(\ref{27})),}%
\]%
\[
\left\langle \varepsilon\right\rangle =E\text{ (Eq. (\ref{2})) and}%
\]%
\[
\sigma_{\varepsilon}=\left\{
\begin{array}
[c]{cc}%
\sigma & E\leq0\\
\sqrt{\frac{2E}{\mu}} & E>0
\end{array}
\right.  \text{ (Eqs. (\ref{22}) and (\ref{7})).}%
\]
If one is not interested in the large or in the small $s$ behavior one may use
the much simpler Eqs. (\ref{smalls}) or (\ref{larges}), respectively, instead
of Eq. (\ref{10}).

Comparing this approximation with the numerically calculated FPT distribution
of the RF model we showed that the first may serve as a good approximation to
the second. Finally, similar methods can be used to approximate other, say
absorbing, boundary conditions at $i=d+1$.

\begin{acknowledgments}
We thank S. Redner for useful discussions. This work was supported by the High
Council for Scientific and Technological Cooperation between France and
Israel. M. S. and Y. K. were also supported by the Israeli Science Foundation,
and O.B. and R.V. by ANR grant "Dyoptri".
\end{acknowledgments}

\bibliographystyle{prsty}
\bibliography{P3}

\end{document}